\documentclass[twocolumn, pra, showpacs,superscriptaddress]{revtex4}
\usepackage{graphicx}
\usepackage{dcolumn}
\usepackage{bm}
\usepackage{amsmath}

\begin{document}

\title{Ground-state properties of one-dimensional anyon gases}
\author{Yajiang Hao}
\affiliation {Beijing National Laboratory for Condensed Matter
Physics, Institute of Physics, Chinese Academy of Sciences, Beijing
100080, P. R. China}
\author{Yunbo Zhang}
\affiliation{Department of Physics and Institute of Theoretical
Physics, Shanxi University, Taiyuan 030006, P. R. China}
\author{Shu Chen}
\email{schen@aphy.iphy.ac.cn} \affiliation{Beijing National
Laboratory for Condensed Matter Physics, Institute of Physics,
Chinese Academy of Sciences, Beijing 100080, P. R. China}
\date{\today}

\begin{abstract}
We investigate the ground state of the one-dimensional interacting
anyonic system based on the exact Bethe ansatz solution for
arbitrary coupling constant ($0\leq c\leq \infty$) and statistics
parameter ($0\leq \kappa \leq \pi$). It is shown that the density
of state in quasi-momentum $k$ space and the ground state energy
are determined by the renormalized coupling constant $c'$. The
effect induced by the statistics parameter $\kappa$ exhibits in
the momentum distribution in two aspects: Besides the effect of
renormalized coupling, the anyonic statistics results in the
nonsymmetric momentum distribution when the statistics parameter
$\kappa$ deviates from $0$ (Bose statistics) and $\pi$ (Fermi
statistics) for any coupling constant $c$. The momentum
distribution evolves from a Bose distribution to a Fermi one as
$\kappa$ varies from 0 to $\pi$.  The asymmetric momentum
distribution comes from the contribution of the imaginary part of
the non-diagonal element of reduced density matrix, which is an
odd function of $\kappa$.  The peak at positive momentum will
shift to negative momentum if $\kappa$ is negative.

\end{abstract}
\maketitle
% 05.30.Pr Fractional statistics systems
% 03.75.Hh Static properties of condensates; thermodynamical, statistical and structural properties.
% 05.30.Fk Fermion systems and electron gas
% 05.30.Jp Boson systems
% 67.40.Db Quantum statistical theory; groundstate, elementary excitations

%\keywords{Bethe ansatz}
%\preprint{APS/123-QED}

\narrowtext
\section{introduction}
The physical systems with fractional statistics have been a
subject of great interest \cite{Wilczek,anyons,Haldane}.
 Although initial studies of fractional statistics
have focused on two-dimensional systems
\cite{anyons,Laughlin,Halperin,Camino}, anyons (the particles
obeying the fractional statistics) have also found application in
various one-dimensional (1D) systems
\cite{Haldane,WangZD,Kundu99,Girardeau06}. Recently, many
theoretical works have been dedicated to study the 1D anyon gas
\cite{Girardeau06,Batchelor06PRL,Patu07,Batchelor07,Batchelor06PRB,Calabrese,anyonTG,
Patu08,Cabra,Patu082,Zhu} since the integrable model of anyon gas
with a $\delta$-function interaction was introduced by Kundu
\cite{Kundu99} despite no proof of the realization of anyon gas in
1D. Because the statistical properties is related with the
topological order, currently anyons also draw intensive attention in
topological quantum computation. Particularly, with the rapid
progress in the regime of cold atom and with the excellent
controllability of the neutral atoms trapped in the optical lattice,
cold atom has become a popular experimental platform in many
research regimes. Rotating Bose-Einstein condensates and cold atoms
in optical lattice have also been proposed to create, manipulate,
and test anyons \cite{PZoller,Aguado,Jiang}.

On the other hand, the experimental progress of trapped 1D cold atom
systems \cite{gorlitz,Paredes,Toshiya,esslinger} has triggered more
and more theorists to study the many body physics of 1D correlated
systems beyond the mean-field theory
\cite{Olshanii,Blume,Petrov,Dunjko,Chen,Olshanii2}. Further, the
ability of tuning the effective 1D interactions by Feshbach
resonance leads experiment accessible to the strong correlation
regime \cite{Paredes,Toshiya}. In the light of recent experiments,
studies of the integrable model have re-attracted much attention
today \cite{Guan05,Sakmann,Hao06,Hao07,Sykes} because with its exact
solution we can obtain the properties in the full physical regime
even in the strong correlated limit. By applying the exact solution
of Lieb-Liniger to the 1D interacting Bose gas in a hard-wall trap
\cite{Lieb,Gaudin}, the evolution from the Bose-Einstein condensate
to ``fermionized" Tonks-Girardeau (TG) gas was explicitly and
exactly displayed in our previous works \cite{Hao06,Hao07}, which
has also been confirmed by various numerical methods
\cite{Zoellner,Deuretzbacher,Landau}. Although the density
distribution of Bose gas in the limit of infinite repulsion is
identical to the free Fermi gas \cite{Girardeau}, the momentum
distributions exhibit quite different properties from the free Fermi
gas because of their respective statistics \cite{Lenard,Vaidya}. It
is well known that the different permutation symmetry leads to the
physics of Bose gases in many aspects differing from the physics of
Fermi gases. As a natural generalization, the anyon gas interpolates
between the Bose and Fermi gas. No doubt the integrable anyon model
provides us the possibility to study the effect of generalized
permutation symmetry in an exact manner although it is still a
theoretically hypothetical model. So far much attention has focused
on its thermodynamic properties and correlation functions
\cite{Batchelor06PRL,Patu07}, however no general result for the
momentum distribution has been given except for the anyon gas in the
infinite interaction limit (or the anyonic TG gas)
\cite{anyonTG,Patu08}. In this paper, we investigate the ground
state of the 1D anyonic system with finite number of anyons in the
whole regime of interaction parameter, and particularly we shall
study how the fractional statistics affect the ground state
properties, such as the ground state energy and the momentum
distribution.  By numerically solving the Bethe ansatz equations, we
obtain the ground state wave function and thus the momentum
distribution for different coupling constant and statistics
parameter. The obvious properties of anyonic statistics are
displayed in the momentum distribution and with the change of
statistics parameter the system exhibits the Bose statistics, Fermi
statistics and the fractional statistics in between.

The paper is organized as follows. In Sec. II, we give a brief
review of 1D anyonic model and introduce its solution with Bethe
ansatz method. In Sec. III, we present the density of ground state
in quasi-momentum space, ground state energy and momentum
distributions for different coupling constant and statistics
parameter. A brief summary is given in Sec. IV.

\section{formulation of the model and its solution}
We consider the one dimensional system of $N$ anyons in length $L$
with second quantized Hamiltonian
\begin{eqnarray}
H_A &=&\frac{\hbar ^2}{2m}\int_0^Ldx\frac{\partial \Psi _A^{\dagger }}{%
\partial x}\frac{\partial \Psi _A}{\partial x} \nonumber \\
&&+\frac {g_{1D}}{2} \int_0^Ldx\Psi _A^{\dagger }\left( x\right)
\Psi _A^{\dagger }\left( x\right) \Psi _A\left( x\right) \Psi
_A\left( x\right) ,
\end{eqnarray}
in which $m$ is the mass of anyon and $g_{1D}$ denotes the effective
interacting constant between anyons. The field operator $\Psi _A^{\dagger
}\left( x\right) $ and $\Psi _A\left( x\right) $ obey the anyonic
commutation relations
\begin{eqnarray*}
\Psi _A\left( x_1\right) \Psi _A^{\dagger }\left( x_2\right)  &=&e^{-i\kappa
\epsilon \left( x_1-x_2\right) }\Psi _A^{\dagger }\left( x_2\right) \Psi
_A\left( x_1\right)  \\
&&+\delta \left( x_1-x_2\right) , \\
\Psi _A^{\dagger }\left( x_1\right) \Psi _A^{\dagger }\left( x_2\right)
&=&e^{i\kappa \epsilon \left( x_1-x_2\right) }\Psi _A^{\dagger }\left(
x_2\right) \Psi _A^{\dagger }\left( x_1\right) ,
\end{eqnarray*}
where the sign function $\epsilon \left( x\right) $ gives $-1$,
$0$ or $1$ depending on whether $x$ is negative, zero, or
positive. $\kappa $ is the parameter related with statistics and
will be restricted in the regime $\left[ 0,\pi \right] $ in the
present work. Particularly, $\kappa = 0$ and $\pi $ correspond to
Bose statistics and Fermi statistics, respectively. A standard
rescaling procedure brings the Hamiltonian into a dimensionless
one
\[
H_A=\int_0^Ldx\left[ \frac{\partial \Psi _A^{\dagger }}{\partial x}\frac{%
\partial \Psi _A}{\partial x}+c\Psi _A^{\dagger }\left( x\right) \Psi
_A^{\dagger }\left( x\right) \Psi _A\left( x\right) \Psi _A\left( x\right)
\right]
\]
with $c$ being the dimensionless coupling constant. The Hamiltonian take the
same form as the Lieb-Liniger Bose model except that the field operator
satisfy anyonic statistics. Similar to the Bose model the eigen problem for
the Hamiltonian can be reduced to solve the problem of $N$-anyons with $%
\delta $-interaction \cite{Kundu99,Batchelor06PRL}
\begin{equation}
H\Psi \left( x_1,\cdots ,x_N\right) =E\Psi \left( x_1,\cdots
,x_N\right)
\end{equation}
with
\begin{equation}
H=-\sum_{j=1}^N\frac{\partial ^2}{\partial x_j^2}+2c\sum_{j<l}\delta
(x_j-x_l).
\end{equation}
The anyonic commutation relations require the wavefunction satisfy the
generalized symmetry
\begin{eqnarray}
&&\Psi \left( x_1,\cdots ,x_j,\cdots ,x_l,\cdots ,x_N\right)   \nonumber \\
&=&e^{-i\omega }\Psi \left( x_1,\cdots ,x_l,\cdots ,x_j,\cdots
,x_N\right) \label{smy}
\end{eqnarray}
with the anyonic phase
\[
\omega =\kappa \left[ \sum_{k=j+1}^l\epsilon
(x_j-x_k)-\sum_{k=j+1}^{l-1}\epsilon (x_l-x_k)\right] .
\]
Using the coordinate Bethe ansatz we can obtain the wavefunction
\begin{eqnarray}
\Psi \left( x_1,\cdots ,x_N\right)  &=&\sum_Q\theta \left(
x_{q_N}-x_{q_{N-1}}\right) \cdots \theta \left( x_{q_2}-x_{q_1}\right)
\nonumber \\
&&\times \exp \left( -i \frac{\kappa} {2}\Lambda \left(
x_{q_1},x_{q_2},\cdots
,x_{q_N}\right) \right)   \nonumber \\
&&\times \varphi _Q\left( x_{q_1},x_{q_2},\cdots ,x_{q_N}\right) ,
\end{eqnarray}
where $Q$ is used to label the region $0\leq x_{q_1}\leq
x_{q_2}\leq \cdots \leq x_{q_N}\leq L$ and $\theta (x-y)$ is the
step function. In different region the phase factors of the
wavefunction are determined by $\Lambda =\sum_{i<j}\epsilon \left(
x_i-x_j\right) $, which results in the exchange symmetry dictated
by eq. (\ref{smy}). The wave function $\varphi_Q \left( x_1,\cdots
,x_N\right) $ is taken as the Bethe ansatz type
\begin{equation}
\varphi_Q \left( x_{q1},\cdots ,x_{qN}\right) =\sum_P\left[
A_{P_1\cdots P_N } \exp \left( i\sum_jk_{p_j}x_{q_j}\right)
\right] , \label{wavfunc}
\end{equation}
In order to get a physical result, the twisted boundary condition
\cite{Patu07}
\[
\Psi  \left( 0,x_2,\cdots ,x_N\right) =e^{i\kappa \left(
N-1\right) }\Psi  \left( L,x_2,\cdots x_N\right)
\]
is used here. Under the twisted boundary condition, we have the
Bethe ansatz equations
\begin{equation}
\exp \left( ik_jL\right) =\prod_{l=1\left( \neq j\right) }^N\frac{%
ik_l-ik_j+c^{\prime }}{ik_l-ik_j-c^{\prime }},  \label{BAE1}
\end{equation}
with $j=1,2,\cdots ,N$, where
\[
 c^{\prime }=c/\cos \left( \kappa /2\right)
\]
is the renormalized coupling constant and  $ \kappa $ the
statistics parameter. It is deserved to notice that the system is
reduced to the Lieb-Liniger Boson model when $\kappa =0$ and the
system is reduced to the free Fermi one when $\kappa=\pi$. The
coefficients $A_P$ take the form of
\[
A_{p_1p_2...p_N}=\in _P\prod_{j<l}^N\left(
ik_{p_l}-ik_{p_j}+c^{\prime }\right) ,
\]
in which $\in _P$ denotes a $+(-)$ sign factor associated with
even (odd) permutations. With the set of quantum number $\left\{
k_j\right\} $ known as quasimomentum the energy eigenvalue and the
total momentum can be formulated as
\begin{equation}
E=\sum_{j=1}^Nk_j^2 \label{Energy}
\end{equation}
and $P=\sum_{j=1}^Nk_j$. Taking the logarithm of Bethe ansatz
equations, we have
\begin{equation}
k_jL=n_j\pi -\sum_{l=1\left( \neq j\right) }^N\left( \arctan \frac{k_j-k_l}{%
c^{\prime }}\right) .  \label{bae}
\end{equation}
In the following only the case of $c>0$ will be investigated. The
set of integer $\left\{ n_j\right\} $ determine an eigenstate and
for the ground state $n_j=j-\left( N+1\right) /2$ ($1\leq j\leq
N$).

\section{properties of the ground state}
By numerically solving the set of transcendental equations
eq.(\ref{bae}) , the quasimomentum $\left\{ k_j\right\} $ and thus
the wave function can be decided exactly. For $c>0$, the set of
$k_i$ is unique and real, therefore the density of state in $k$
space $\rho (k)$ can be formulated as \cite{Yang}
\begin{equation}
L\rho \left( \frac{k_j+k_{j+1}}2\right) =\frac 1{k_{j+1}-k_j}.
\label{d_k}
\end{equation}
In Fig. 1, we plot the density of ground state in quasi-momentum
$k$ space with the interaction parameter $c$ fixed to $c=0.1$ for
several $\kappa$. It is shown that the density of state of anyons
with coupling constant $c$ and statistical parameter $\kappa$ is
identical to the density of state of Lieb-Liniger Bose model with
effective coupling constant $c'=c/cos(\kappa/2)$. This is clear by
comparing the Bethe ansatz equation (\ref{bae}) with that of
Lieb-Liniger model \cite{Lieb}. When $\kappa =0$, the model is
reduced to Bose model and $k$ mainly distribute in the regime
around zero. While for the case of $\kappa =\pi$, we have
$k_j=(j-(N+1)/2)2\pi/L$ with $j=1,\cdots, N$ even for the weak
coupling $c=0.1$, which is the same as the situation of
$c'=\infty$ of the Lieb-Liniger Bose model. Here the statistics
parameter $\kappa$ affects the density of state in quasi-momentum
space only by renormalizing the coupling constant $c$ into $c'$.
For the case of $c=0.1$, we note that when $\kappa$ approaches
$\pi$ minor change of $\kappa$ will result in the great change of
the density of ground state in $k$ space. For instance, there are
obvious difference among the densities corresponding to
$\kappa=0.9\pi, 0.99\pi$, and $\pi$ due to the dramatic changes of
the effective coupling constant $c'=0.64, 6.37$ and $\infty$
correspondingly. In Fig. 2, we display the ground-state energy
versus $\kappa$ for different coupling constant $c$. The
ground-state energy increases with the increase of $\kappa$ and
the coupling constant $c$. For the weak coupling the energy of the
system is small for most $\kappa$ but it abruptly increases and
arrives at the maximum when $\kappa$ approach $\pi$. In the strong
coupling limit ($c\rightarrow \infty$) the ground-state energy is
almost a constant independent of the statistics parameter.
Similarly the ground-state energy of the anyon gas with coupling
constant $c$ and statistical parameter $\kappa$ is identical to
that of the corresponding Bose gas with effective coupling
constant $c'=c/cos(\kappa/2)$.

\begin{figure}[tbp]
\includegraphics[width=3.5in]{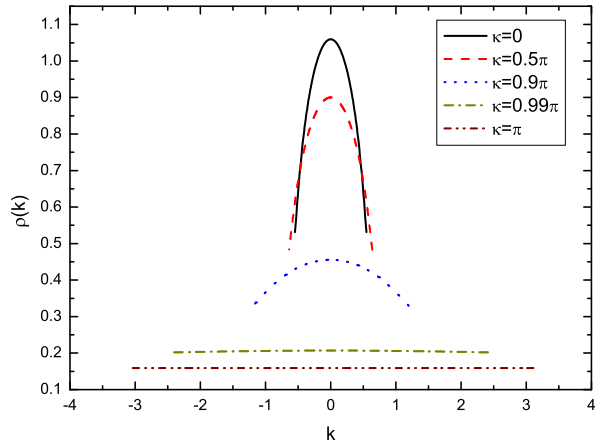}
\caption{ (color online) The density of state in quasi-momentum
$k$ space for the ground state for $N=200$, $L=200$ and $c=0.1$.
$\kappa=0$ (solid line), $\kappa=0.5\pi$ (dashed lines),
$\kappa=0.9\pi$ (dot lines), $\kappa=0.99\pi$ (dash dot lines) and
$\kappa=\pi$ (dash dot dot lines).} \label{fig1}
\end{figure}
\begin{figure}[tbp]
\includegraphics[width=3.5in]{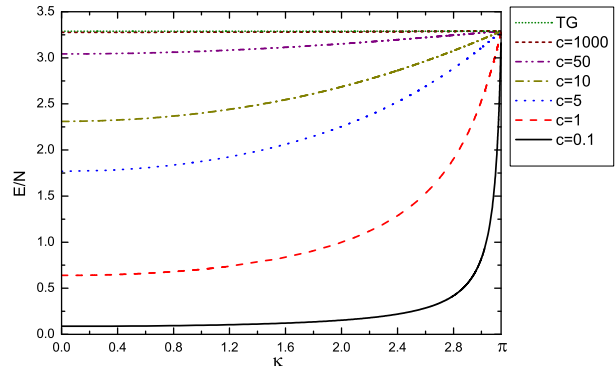}
\caption{(color online) The energy of ground state for $N=200$ and
$L=200$. $c=0.1$ (solid line), $c=1$ (dashed lines), $c=5$ (dot
lines), $c=10$ (dash dot lines), $c=50$ (dash dot dot lines),
$c=1000$ (short dash lines) and $c=\infty$ (short dot lines).}
\label{fig2}
\end{figure}

Now it is clear that the density of state in $k$ space and the
ground-state energy are determined completely by the renormalized
coupling constant $c'$. Furthermore, from the eqs. (\ref{Energy})
and (\ref{bae}), we see that the energy level structure of the
anyon gas is exactly the same as the corresponding bosonic model
with renormalized coupling constant  $c'$. Therefore the
thermodynamic properties of the anyon gas are the same as the well
known thermodynamic properties of the 1D boson gas
\cite{Lieb,Yang} with the effective coupling $c'=c/cos(\kappa/2)$.
This implies that the intrinsic difference between the anyon gas
and the corresponding Bose gas is hard to be distinguished just by
the thermodynamic properties. However, due to the different
exchange symmetry of the wave functions, the observables
associated with the wave functions rather than the square of wave
functions, such as the off-diagonal reduced density matrix and the
momentum distributions, should display quite different behaviors.
In terms of the ground state wave function $\Psi \left( x_1,\cdots
,x_N\right) $, the one-body reduced density matrix is given by
\begin{eqnarray*}
&&\rho (x,x^{\prime }) \\
&=&\frac{N\int_0^Ldx_2\cdots dx_N\Psi ^{*}\left( x,x_2,\cdots
,x_N\right) \Psi \left( x^{\prime },x_2,\cdots ,x_N\right)
}{\int_0^Ldx_1\cdots dx_N\left| \Psi \left( x_1,x_2,\cdots
,x_N\right) \right| ^2},
\end{eqnarray*}
while the momentum distribution
\begin{equation}
n\left( p\right) =\frac 1{2\pi }\int_0^Ldx\int_0^Ldx^{\prime }\rho
(x,x^{\prime })e^{-ip\left( x-x^{\prime }\right) }
\end{equation}
is the Fourier transformation of $ \rho (x,x^{\prime })$. Below we
rescale the momentum distribution $n(p)$ and momentum $p$ into
dimensionless forms in the units of $L/2\pi $ and $2\pi /L $
respectively and the original notations are reserved.
\begin{figure}[tbp]
\includegraphics[width=3.5in]{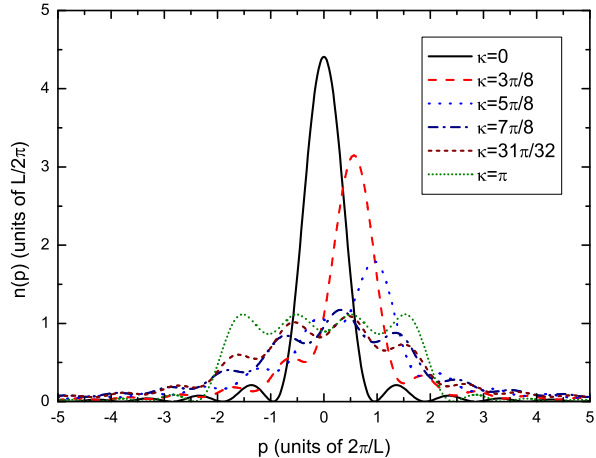}
\caption{(color online) Momentum distribution of anyons for $N=4$,
$L=1$ and $c=0.1$. $\kappa=0$ (solid line), $\kappa=3\pi/8$ (dashed
lines), $\kappa=5\pi/8$ (dot lines), $\kappa=7\pi/8$ (dash dot
lines), $\kappa=31\pi/32$ (short dash lines) and $\kappa=\pi$ (short
dot lines).} \label{fig3}
\end{figure}
\begin{figure}[tbp]
\includegraphics[width=3.5in]{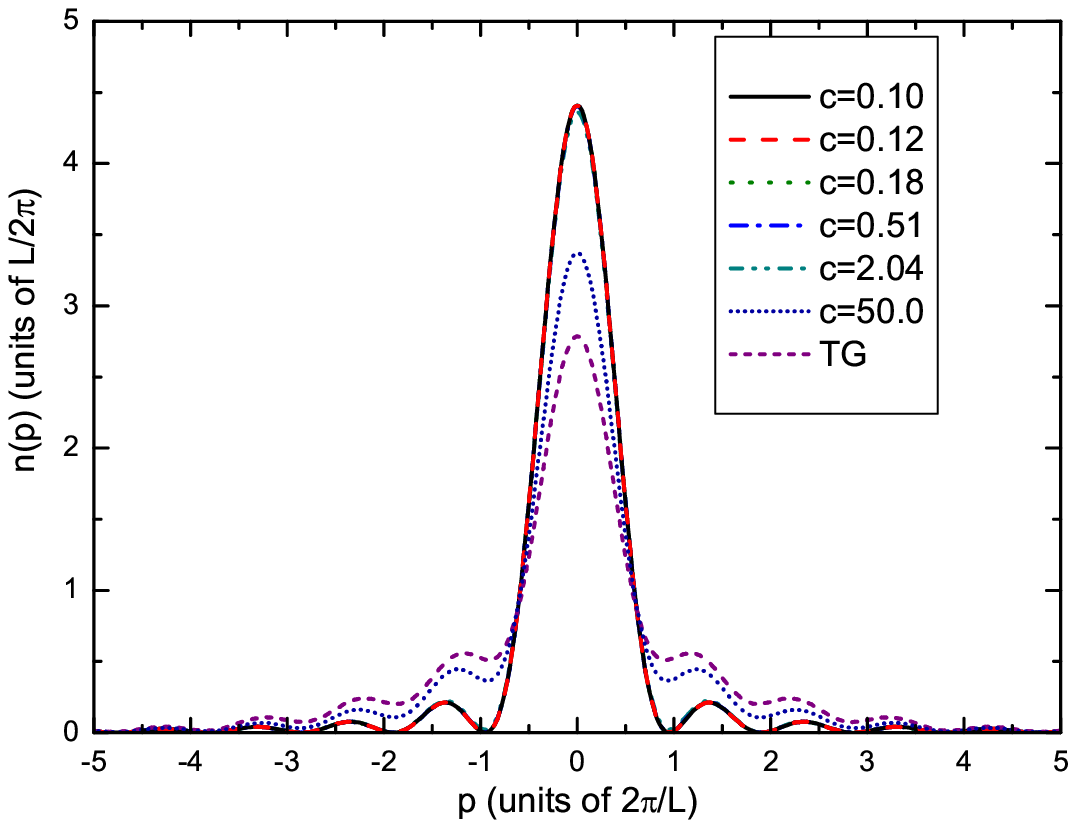}
\caption{(color online) Momentum distribution of Bosons for $N=4$,
$L=1$ and $c=0.1$/cos($\kappa$/2). $\kappa=0$ (solid line),
$\kappa=3\pi/8$ (dashed lines), $\kappa=5\pi/8$ (dot lines),
$\kappa=7\pi/8$ (dash dot lines), $\kappa=31\pi/32$ (short dash
lines) and $\kappa=\pi$ (short dot lines).} \label{fig4}
\end{figure}
\begin{figure}[tbp]
\includegraphics[width=3.5in]{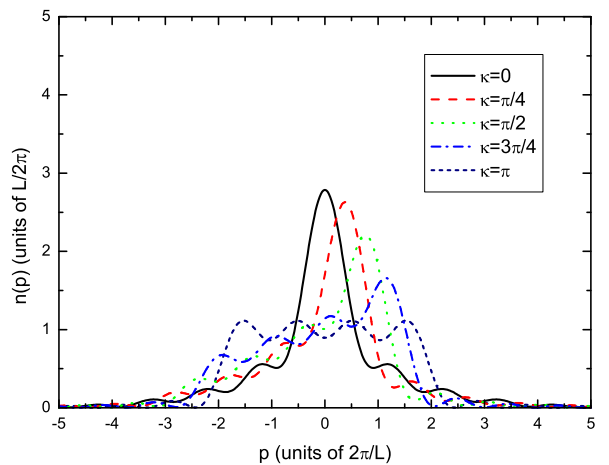}
\caption{(color online) Momentum distribution of anyons for $N=4$,
$L=1$ and $c=\infty$. $\kappa=0$ (solid line), $\kappa=\pi/4$
(dashed lines), $\kappa=\pi/2$ (dot lines), $\kappa=3\pi/4$ (dash
dot lines) and $\kappa=\pi$ (short dash lines).} \label{fig5}
\end{figure}

To give a concrete example, we display the momentum distribution
of the anyon gas  with coupling constant $c=0.1$ and $N=4$ in Fig.
3 for various statistics parameters. At $\kappa =0$, the momentum
distribution shows a typical Boson-type distribution with an
obvious peak around the zero momentum point, whereas the momentum
distribution is identical to that of free spinless Fermions at
$\kappa =\pi$, exhibiting shell structure. In these two situations
the momentum distributions are symmetric. When $\kappa$ deviates
from these two extrema points, the momentum distribution becomes
asymmetric: the anyons distribute with more probability in the
regime of positive momentum whereas with relatively small
probability in the other side. With the increase of $\kappa$, the
distribution redistributes between the regime of positive and
negative $p$, and gets wider with more and more peaks appearing.
Finally, $N$ peaks appear at $\kappa=\pi$ and the momentum
distribution shows the characteristic shell structure of Fermions.

As a comparison, in Fig. 4 we display the momentum distribution of
the Lieb-Liniger boson gas for $N=4$ with the renormalized
coupling constants $c=0.1/cos(\kappa/2)=0.1, 0.12, 0.18, 0.51,
2.04$ and $\infty$ corresponding to those for anyons in Fig. 3,
where we have $c=0.1$ and $\kappa=0, 3\pi/8, 5\pi/8, 7\pi/8,
31\pi/32$, and $\pi$ respectively. The momentum distributions for
$c=0.1, 0.12, 0.18, 0.51, 2.04$ are not obviously different in
Fig. 4 as they all fall in the weakly interacting regime. We also
display the momentum distribution for $c=50$ in Fig. 4.  The
height of the peak shrinks with the increase of the repulsive
interaction and stronger repulsive interaction between the bosonic
atoms tends to spread out the distribution widely. Even in the TG
limit, the momentum distribution of bosons does not show shell
structure like free fermions.  For all coupling constant, bosons
always congregate symmetrically in the regime around $p=0$ with
the most probability and with the increase of momentum $p$ the
probability diminishes rapidly. Obviously, the asymmetry in the
momentum distribution of anyons can not be attributed to
renormalization of the effective coupling constant and is induced
completely by the the permutation symmetry related to statistic
parameter $\kappa$, which can be regarded as the indication of
anyons.

In order to clarify more clearly the effect induced by the
statistic parameter $\kappa$, we show the momentum distributions
for the system with $N=4$ and $c=\infty$ in Fig. 5. In this case
the renormalized coupling constant $c'=\infty$ and the
quasimomentum $\left\{k_j \right\} $ shall be same for arbitrary
$\kappa$, therefore the statistics parameter only results in
different permutation symmetry embodied in the wavefunction. There
appears the similar situation as the case for $c=0.1$ in Fig. 3.
In the Bose limit ($\kappa =0$) and in the Fermi limit ($\kappa =
\pi$) the momentum distributions are symmetric while in between
the distributions take on the nonsymmetric profiles and evolve
from a Bose distribution into a free spinless Fermi distribution
with the increase of $\kappa$. The difference from the case of
weak coupling is that the system distributes in a wider regime
relative to that of weak coupling.  Mathematically, the asymmetric
momentum distribution is induced by the existence of imaginary
part of the density matrix (see appendix for detail), which is an
odd function of statistical parameter such that the peak at
positive momentum in Fig. 3 and Fig. 5 is a result of the choice
of positive $\kappa$, and the peak will shift to negative momentum
if $\kappa$ is negative.

\section{conclusions}
In summary, we have investigated the ground-state properties of 1D
anyon gas based on the exact Bethe ansatz solution for arbitrary
coupling constant ($0\leq c\leq \infty$) and statistics parameter
($0\leq \kappa \leq \pi$). The density of state in quasimomentum
space and the ground-state energy are determined by the
renormalized coupling constant $c'$. The anyonic system with
coupling constant $c$ and statistics parameter $\kappa$ has the
same density of state and energy as those of the Lieb-Liniger Bose
model with coupling $c'=c/cos(\kappa$/2). Besides the
renormalization of the effective coupling constant, the additional
effect induced by the statistics exhibits in the momentum
distribution. While in the limit of $\kappa =0$ and $\kappa =\pi$,
the anyon gas is reduced to the Bose gas and free spinless Fermi
gas, respectively, with both of momentum distributions being
symmetric. When the statistics parameter deviates from these two
points, the momentum distribution is asymmetric and evolves from
the Bose distribution to Fermi distribution with the increase of
$\kappa$. Although the thermodynamic properties of anyon sytem are
even functions of the statistic parameter $\kappa$ because the
renormalized coupling constant is even function of it, the
momentum distribution is not so because of the existence of the
imaginary part of the non-diagonal part of the one body density
matrix, which is an odd function of $\kappa$. If $\kappa$ is
negative the momentum distribution is equal to the distribution
for positive $\kappa$ with the mapping between the positive
momentum and negative momentum. Our results interpolate between
the known results of Bose and Fermi gas and clarify the effect of
generalized permutation symmetry of the anyon gas in an exact
manner.

\begin{acknowledgments}
S.C. would like to appreciate X. W. Guan for useful discussions.
This work is supported by NSF of China under Grants  No. 10574150
and No. 10774095, National Program for Basic Research of MOST
China, the 973 Program under Grant No. 2006CB921102, and Shanxi
Province Youth Science Foundation under Grant No. 20051001.

\end{acknowledgments}

\appendix
\section{}

The asymmetry of the momentum distribution can be attributed to
the special permutation symmetry of the anyon wavefunction
characterized by the statistical parameter $\kappa $. To see it
clearly, we display the case of two anyons in detail in this
appendix. For the two anyon systems, the Bethe ansatz wavefunction
takes the form of
\begin{eqnarray*}
\Psi \left( x_1,x_2\right)  &=&\theta \left( x_1<x_2\right) \exp
\left(
i\kappa /2\right) \varphi \left( x_1,x_2\right) + \\
&&\theta \left( x_2<x_1\right) \exp \left( -i\kappa /2\right)
\varphi \left( x_2,x_1\right) ,
\end{eqnarray*}
where $ \varphi \left( x_1,x_2\right)= A_{12} \exp(ik_1 x_1 + i
k_2 x_2) + A_{21} \exp(ik_2 x_1 + i k_1 x_2) .$ The reduced one
body density matrix $\rho \left( x,x^{\prime }\right)= \int_0^L
dx_2\Psi ^{*}\left( x,x_2\right) \Psi \left( x^{\prime
},x_2\right)$ can be evaluated with wavefunction, i.e.,
\begin{eqnarray*}
&&\rho \left( x,x^{\prime }\right)  \\
&=&\int_0^Ldx_2\left[ \theta \left( x<x_2\right) \theta \left(
x^{\prime }<x_2\right) \varphi ^{*}\left( x,x_2\right) \varphi
\left( x^{\prime
},x_2\right) \right.  \\
&&+\theta \left( x<x_2\right) \theta \left( x_2<x^{\prime }\right)
\exp \left( -i\kappa \right) \varphi ^{*}\left( x,x_2\right) \varphi
\left(
x_2,x^{\prime }\right)  \\
&&+\theta \left( x_2<x\right) \theta \left( x^{\prime }<x_2\right)
\exp \left( i\kappa \right) \varphi ^{*}\left( x_2,x\right) \varphi
\left(
x^{\prime },x_2\right)  \\
&&\left. +\theta \left( x_2<x\right) \theta \left( x_2<x^{\prime
}\right) \varphi ^{*}\left( x_2,x\right) \varphi \left(
x_2,x^{\prime }\right) \right] .
\end{eqnarray*}
When the statistical parameter $\kappa $ deviates from $0$ and
$\pi $, the density matrix is always complex for $x \neq x^{\prime
}$. From the definition of reduced density matrix, we have $ \rho
(x,x^{\prime })=\rho ^{*}(x^{\prime },x)$, which implies
$\mathop{\rm Re} \left[ \rho (x,x^{\prime })\right]=\mathop{\rm
Re} \left[ \rho (x^{\prime },x)\right]$ and $\mathop{\rm Im}
\left[ \rho (x,x^{\prime })\right]=-\mathop{\rm Im} \left[ \rho
(x^{\prime },x)\right]$. Therefore the momentum distribution can
be represented as
\begin{eqnarray*}
n\left( p\right)  &=&\frac 1{2\pi }\int_0^Ldx\int_0^Ldx^{\prime
}\rho
(x,x^{\prime })e^{-ip\left( x-x^{\prime }\right) } \\
&=&\frac 1{2\pi }\int_0^Ldx\int_0^Ldx^{\prime }\left\{ \mathop{\rm
Re} \left[ \rho (x,x^{\prime })\right] \cos p\left( x-x^{\prime
}\right) \right.
\\
&&\left. + \mathop{\rm Im} \left[ \rho (x,x^{\prime })\right] \sin
p\left( x-x^{\prime }\right) \right\} .
\end{eqnarray*}
The second term is the odd function of momentum $p$ such that the
momentum distribution becomes asymmetric about $p$. Furthermore,
the imaginary part $\mathop{\rm Im} \left[ \rho (x,x^{\prime
})\right]$ is an odd function of $\kappa$. Therefore, if we take
$\kappa$ as negative ($\kappa \rightarrow - \kappa$), the peak at
positive momentum as shown in Fig.3 and Fig.5 will shift to
negative momentum.

\end{document}